# SIMULATED EFFECTS OF ODD-ALKANE IMPURITIES IN A HEXANE MONOLAYER ON GRAPHITE

Cary L. Pint* and M.W. Roth

Department of Physics, University of Northern Iowa, Cedar Falls, IA  50614-0150
*electronic mail:  cpint@uni.edu

## ABSTRACT

We present the results of molecular dynamics (MD) simulations of odd alkane impurities present within the hexane (even alkane) monolayer.  We simulate various temperatures at ca. 3%, 5%, 10%, and 15% impurities of propane ($C_3H_8$), pentane ($C_5H_{12}$), heptane ($C_7H_{16}$), nonane ($C_9H_{20}$), and undecane ($C_{11}H_{24}$), each having a low-temperature solid phase belonging to a different space group as compared to hexane, to study the effects of impurities on the various phases and phase transitions for hexane monolayers that are well-characterized through previous experimental and theoretical work.  Based upon preferential adsorption, we provide two emerging pictures of how impurities could affect the monolayer, for impurity chain lengths longer and shorter than that of the hexane molecules.  We provide evidence that impurities in the monolayer, even in small proportion to the hexane, could induce significant changes in the phase behavior and phase transitions, and we propose that because of the size of pentane with respect to hexane, and the nature of the solid herringbone (HB) phase, pentane impurities give the best representation of the phase behavior observed for the pure hexane monolayer.  We find that impurities with chain lengths longer than hexane tend to distort the sublattices of the solid HB phase, and hence lead to a phase transition into an "intermediate" phase significantly prior to that observed for a pure hexane monolayer.  We discuss an interpretation of this study that could be potentially useful for future experimental work.

# I. INTRODUCTION

The study of molecules adsorbed onto surfaces has been a topic of considerable interest in recent years due to advances in methods of studying these systems experimentally as well as the theoretical techniques and increasing computer performance that allows simulations to be an excellent tool for investigation into these systems as well. Of particular interest in such studies has been the family of linear short-chained *n*-alkanes, whose unique combination between a molecule with a linear chain length (and hence significant number of degrees of freedom) and properties that are still extremely useful for industrial applications (such as lubrication, adhesion, wetting on the surface, etc.) has motivated several studies of adsorption of monolayers and multilayers of these alkanes physisorbed onto the surface.

Since the *n*-alkanes are very abundant in nature in multi-component mixtures (such as various petroleum products), their usefulness in industrial applications is usually limited to primarily those applications that involve multi-component mixtures, as these are much cheaper and more readily available than pure alkanes, which usually tend to be the focus of experimental studies (due to the complex adsorption behavior of mixed alkane solutions). Therefore, an understanding of how the dynamics and energetics of the phases and the phase transitions in mixtures differ from those in pure monolayers is of fundamental interest to both of these.

In this study, we investigate the presence of impurities in an even alkane (hexane) monolayer. The hexane monolayer is specifically chosen because due to the significant amount of both experimental and theoretical work that has been previously completed regarding the study of the phases and phase behavior the monolayer exhibits. Previous experimental work completed over hexane monolayers[1-3] has indicated that the monolayer arranges in a solid herringbone (HB) phase at low temperatures. This phase then continues until $T$=151K where experiment reports that commensurability is lost with the substrate and the phase behavior involves what diffraction patterns suggest to be rectangular-centered (RC) islands coexisting with a liquid phase. This persists until ca. 176K where a melting transition takes place into an isotropic fluid. Further experimental work[4] over hexane monolayers and multilayers, as well as several other short-chained even alkanes, reports the same type of behavior, and reports that for multilayers of short-chained *n*-alkanes, the solid-liquid coexistence region corresponds to what these authors refer to as a solid monolayer persisting above the bulk melting point.

Simulations conducted in recent years[5-11] have confirmed many of these experimental observations. In particular, most recent simulations[11] of hexane monolayers indicate the presence of an "intermediate" phase with nematic order (or RC order) that exists between the solid HB phase and the fluid phase. However, simulations predict this transition (into the intermediate phase) occurs at ca. 138K, and it is believed that the absence of a coexistence region in simulations is largely due to the finite size of systems available to simulations for study.

The impurities that are studied in the hexane monolayer involve chain lengths that vary in length from propane ($C_3H_8$) to undecane ($C_{11}H_{24}$). With the exception of propane, whose low temperature phase behavior is not yet fully understood[12], the odd-alkanes studied occupy a different 2D space group[13] (*cm*, involving a low-temperature rectangular-centered solid phase) than the even alkanes[4] (*pgg*, involving a low-temperature solid HB phase). Propane monolayers[12] are suggested to exhibit a low-temperature phase that most closely resembles a HB phase, but recent diffraction experiment has not been able to verify this.

When dealing with such impurities, where there is a chain-length mismatch between the impurity molecules and the hexane molecules (for this specific case), one must consider the theory of preferential adsorption first proposed by Groszek[14] for linear chain *n*-alkanes. This theory proposes that the longer-chained alkane component in a binary mixture will tend to adsorb onto the surface first, due to an increased number of hydrogen atom "contacts" per unit area. Recent experimental work[15-18] has been completed studying preferential adsorption in *n*-alkane solutions with components of similar size. Based on preferential adsorption, experimental studies of octane and nonane mixtures at ca. 0.8 monolayers indicates that adsorption of octane into the nonane solid phase and that until ca. *X*=0.53 (where *X* indicates the mole fraction of octane) the monolayer forms a molecular compound on the surface. Most recent work[18] studies the surface freezing (SF) effect in binary *n*-alkane mixtures where the difference in the length of the alkanes in the mixture is both small and significant. In the case where the length difference is small, these authors observe that the solid monolayer formed above the bulk melting temperature involves a homogenous mixture of the two alkane components. For a large length difference, these authors report that the monolayer is more richly inhibited by the longer alkane component, consistent with previous observations of preferential adsorption.

At this point, the reader should be cautioned into a close comparison of impurities in the monolayer as being in direct relation to studies of *mixing* behavior for alkane mixtures where the mole fractions of the two components are very similar. In fact, to give the most reliable interpretation of what impurities in the monolayer mean, one generally has to consider the two cases of impurities with chain lengths (*i*) smaller than and (*ii*) greater than that of the pure hexane monolayer, in which they are adsorbed. In the first case, with impurities present in the monolayer with chain lengths smaller than the hexane molecules, the effect of impurities could be due to a small mole fraction of the shorter alkane adsorbed from a binary solution of the shorter alkane/hexane where the hexane molecules would be preferentially adsorbed onto the surface (and thus would determine the solid HB that the shorter alkanes would be adsorbed into). The second case, where the impurity chain length is longer than the hexane molecules, would be representative of *defects* in the monolayer as a result of an impure solution. In fact, most solutions that are utilized for experimental studies of pure alkanes involve solutions that are ca. 96-99% pure. In such cases, the presence of even a small amount of a longer chain alkane (which, by preferential adsorption will tend to adsorb onto the surface prior to the hexane molecules) could create crystallization defects in the monolayer, which could largely affect the phase behavior observed in experiment. It is well known that crystallization on a surface is never, by any means, a "perfect" process, and often involves several types of defects that can dramatically affect the behavior of the crystalline monolayer that is formed.

In addition to preferential adsorption, phase separation between the two components in the monolayer with mismatched chain lengths is also an issue. Recent experimental work[19] has been completed which studies the microphase separation of metastable paraffin mixtures. In general, this work reports the relaxation that occurs involving the system migrating into an equilibrium state depends both on the chain length mismatch and the relative composition of the two paraffin types in the mixture. Although, the time scales over which this relaxation occurs is extremely long compared to the time scales that MD simulations can capture (on the order of $10^3$ minutes), thus the effects of such separation can be effectively neglected in our simulations. However, the layer promotion that tends to ensue in the intermediate phase for hexane monolayers as reported in recent simulations[11] gives an interesting medium by which aspects of preferential adsorption and phase separation can be studied, as will be discussed later.

Therefore, the scientific motivation of this work is to utilize a monolayer that has been well-characterized both experimentally and theoretically, and study how odd-alkanes, whose low-temperature phase behavior has been studied and found to be significantly different than that of the even-alkanes (and hence, hexane) monolayer, affect the monolayer when present as impurities. In particular, based upon the chain-length effects of impurities shorter and longer than the hexane molecules, the goal of this work becomes two-fold. The first case is to study adsorption of hexane from multi-component mixtures of hexane and a shorter alkane solution, where hexane will be expected to preferentially adsorb in greater concentration and in a solid HB phase in the monolayer based upon preferential adsorption. The second case involves impurities with chain lengths longer than hexane, where preferential adsorption would govern that the longer chain molecules adsorb onto the surface prior to the hexanes, presenting a situation arising with impurities present in the monolayer in greater abundance than there were initially present in the solution. Previous experimental work that has studied both odd and even alkane monolayers has concluded the following about the differing solid phase behavior in such monolayers, "This odd-even variation will be expected to have profound consequences in the mixing behavior of odd and even alkanes on the surface." This work is study if there truly are "profound consequences" introduced by the presence of impurity defects in the solid monolayer.

## II. POTENTIAL MODEL

The potential model that is utilized in this work has been described in detail in previous work[11], so only a brief description will be given here. In short, the potential model consists of both bonded and non-bonded interactions. The two considered non-bonded interactions are those of the adatom-adatom interaction and the graphite-adatom interaction. To simulate the adatom-adatom interaction, we utilize a 12-6 Lennard-Jones pair potential function with interaction parameters fit for a psuedoatom representation of the *n*-alkanes. To simulate the interaction between the graphite surface and the psuedoatoms, a Fourier expansion proposed by W.A. Steele[20], that represents the interaction between a psuedoatom and an infinite number of graphene planes is utilized.

The bonded interactions are in the form of bond-angle bending and dihedral angle (torsional) bending. In the three-body bonded interaction (bond-angle bending), the bond-angles are assumed to be harmonic and the pair potential proposed in [21] is utilized. To represent the torsion in the molecular chains, a potential given in [22] is used. Both of these have been utilized in previous work in which good experimental comparison is reported. In all simulations, the RATTLE[23] algorithm is utilized to constrain the solution to the equations of motion in such a way to keep the bond lengths at their equilibrium values of 1.54 Å. For more information regarding simulation parameters and the mathematical forms of the potential functions utilized, the reader is referred to [11].

## III. SIMULATION DETAILS

In all simulations, a constant particle number, constant planar density, and constant temperature canonical ensemble ($N$, $\rho$, $T$) molecular dynamics (MD) method is utilized to carry out atomistic simulations of hexane with 3%, 5%, 10%, and 15% impurities of differing odd alkanes occupying spaces within the hexane monolayer. The united atom (UA) model is used to model each molecule, which approximates each methyl ($CH_3$) and methylene ($CH_2$) group as a single pseudoatom with an increased Van Der Waals radius from that of a single carbon atom. The UA model has been widely used in previous simulations[5-11] and previous work[8] that studies the difference between the UA and AUA models for hexane monolayers indicates little difference in the phase behavior observed, with the only differences referring to primarily the amount of rolling and tilting of the molecules with respect to the substrate, both of which do not seem to affect the phase transitions significantly.

For each simulation of differing numbers of impurity molecules, the total number of pseudoatoms decreases or increases depending upon whether the chain length of the impurity molecule is smaller or larger than the chain length of hexane. However, in all simulations, 112 molecules is used to occupy a cell size of $a$=68.1735Å and $b$=68.88Å, where $a$ and $b$ correspond to orthogonal axes parallel to the plane of the substrate, and the cell is kept as square as possible to minimize the effects of periodic boundary conditions. Therefore, with a total of 112 molecules, the total number of impurity molecules that are simulated are 3, 6, 11, and 17, corresponding to 2.68%, 5.36%, 9.82%, and 15.18%. However, for all practical purposes, these will be referred to 3%, 5%, 10%, and 15% respectively. Although this method does not conserve area per psuedoatom, it does conserve area per molecule, which keeps a rigid solid herringbone phase, to alleviate deviation from the coverage of a complete monolayer. The impurities that are simulated are propane (C3), pentane (C5), heptane (C7), nonane (C9), and undecane (C11). For each impurity, a temperature sequence of $100 \leq T \leq 200$ consisting of 18 points with the points nearest transitions and in the temperature region of the intermediate nematic-ordered phase (previously observed for monolayer hexane) incremented in steps of 5K, for each of 3%, 5%, 10%, and 15% of that respective impurity. This is then repeated for all impurities that are studied.

In all simulations, periodic boundary conditions are used in the $a$ and $b$ directions (where the $a$-axis corresponds to the conventional $x$-axis direction, and the $b$-axis corresponds to the conventional $y$-axis direction, with respect to the computational cell), with free boundary conditions imposed to the plane perpendicular to the substrate. To achieve temperature control, the velocities are frequently rescaled to satisfy equipartition for the center-of-mass, rotational, and internal temperatures:

$$T_{CM} = \frac{1}{3N_m k_B} \sum_{i=1}^{N_m} M_i v_{i,CM}^2, \qquad (1)$$

$$T_{ROT} = \frac{1}{3N_m k_B} \sum_{i=1}^{N_m} \omega_i^T I_i \omega_i, \qquad (2)$$

$$T_{INT} = \frac{1}{2n_C - 5} \sum_{i=1}^{N_m} \sum_{j=1}^{n_C} m_{ij} (v_{ij} - v_{i,CM} - \omega_{i,CM} \times r_{ij,CM})^2. \qquad (3)$$

Here $T_{CM}$, $T_{ROT}$ and $T_{INT}$ are the respective temperatures of the system and the summation index *(i)* runs over molecules and the index *(j)* runs over pseudoatoms within a given molecule. The method of temperature rescaling is applicable for simulations that achieve an equilibrium state, as our simulations do.

The time step for all simulations is 1 fs, and integration of the equations of motion is completed through use of a velocity Verlet RATTLE algorithm. All simulations in this study are run through a period of 2 x $10^5$ equilibration steps (200 ps), followed by a period of 5 x $10^5$ dynamical steps (500 ps) where important quantities being monitored throughout the simulation are averaged. These long simulation times are found to be necessary to study the phase behavior that is exhibited in the intermediate phase, that exists prior to melting.[11]

## IV. RESULTS

To monitor the system structure and phase behavior, a set of order parameters that classify the different phases and their respective orientational and translational behavior are used. The first of these, *OPher*, is a monitor of the low-temperature HB phase and is defined as:

$$OPher = \frac{1}{N_m} \left\langle \sum_{i=1}^{N_m} \sin(2\phi_i)(-1)^j \right\rangle.$$

This order parameter is a sum over all molecules in the system that is maximized (*OPher*=1) when the molecular orientation about the long-axis of each molecule with respect to the *a*-axis of the computational cell, $\{\phi_i\}$, is $45^0$ or $135^0$. The integer *j* is defined due to the difference of 90° that exists between two consecutive sublattices in the HB phase. Since the glide-line of the graphite substrate has angles of 30° and 150° with respect to the *a*-axis, $\phi_i$ tends to be oriented along the glide-line in the commensurate HB phase giving a value of *OPher*=0.86. Figure 3, figure 4, and figure 7 contain plots of *OPher* for monolayers with 3% and 5% impurities.

The next order parameter, *OPnem*, is a measure of the orientation of the long axis of each molecule about a single angle with respect to the *a*-axis, and is defined as:

$$OPnem = \frac{1}{N_m} \sum_{i=1}^{N_m} \langle \cos 2(\phi_i - \phi_{dir}) \rangle,$$

where $\phi_i$ is the same angle introduced previously for *OPher* and

$$\phi_{dir} = \frac{1}{2} \tan^{-1} \left[ \frac{\sum_{i=1}^{N_m} \sin(2\phi_i)}{\sum_{i=1}^{N_m} \cos(2\phi_i)} \right].$$

In a HB phase, $\phi_{dir} = 0$ and *OPnem* takes on a single value corresponding to the angle $\phi_i$. However, in the case that $\phi_i$ assumes a single orientation parallel to the *a*-axis such that $\phi_i$=0, then *OPnem*=1. Although, as the molecular orientations approach a single orientation along the *a*-axis in such a way, the value of *OPnem* is heightened from single value exhibited in the HB. In the case of a fluid, where $\phi_i$ is randomly sampled over $[0, 2\pi]$, the value of *OPnem* will vanish.

In addition, *OPcom*, measures the molecular center-of-mass commensurability of each molecule and is defined as:

$$OPcom = \frac{1}{3N_m} \sum_{i=1}^{N_m} \left\langle \sum_{s=1}^{6} \exp(-ig_s \cdot r_i) \right\rangle,$$

where the outer sum runs over each of the positions of the molecules, $r_i$, and the inner sum runs over all six graphite reciprocal lattice vectors $g_s$. The value of *OPcom* is maximum when all molecular center-of-masses are positioned directly over graphite hexagon centers, and disappears when molecular positions assume random positions (i.e. in the isotropic fluid) along the plane parallel to the substrate. Although *OPcom* is a very effective order parameter to give valuable information regarding molecular commensurability, another measure of commensurability, but from an *atomistic* point of view is given by the average corrugation energy, $<U_I>$, which is defined as

$$<U_I> = \frac{1}{N_m}\left\langle \sum_{i=1}^{N} E_{ni}(r_i) \right\rangle,$$

where $E_n(r)$ is the potential energy between an adatom $i$ and the corrugation in the graphite substrate. Similarly to *OPcom*, this quantity takes on a value very close to zero when the atomistic positions are randomly sampled in the plane parallel to the substrate. However, when the psuedoatoms are all arranged over hexagon centers, the value of $<U_I>$ becomes very large, indicating that the molecules are atomistically bound to the substrate and the surface interaction is the dominant interaction. *OPcom* and $<U_I>$ are excellent indicators of phase transitions where translational order is lost with respect to the substrate.

Among other energies that were averaged throughout the simulations, the only one with interesting features that gives information about phase behavior and transitions is the average Lennard-Jones energy, $<U_{LJ}>$, and is defined as

$$<U_{LJ}> = \frac{1}{N_m}\left\langle \sum_{i=1}^{N}\sum_{j=i+1}^{N} u_{LJ}(r_{ij}) \right\rangle,$$

where $u_{LJ}(r)$ is the pair interaction energy between adatom $i$ and $j$. This quantity is a very good indicator of the commensurate-incommensurate transition (CIT) where translational order is lost with respect to the substrate and the molecule-molecule interaction becomes dominant. Figure 2, figure 5, and figure 6 all show the temperature dependence of *OPcom*, *OPnem*, $<U_I>$, and $<U_{LJ}>$ for 3% and 5% impurities in the hexane monolayer. In all cases (including those plots of *OPher*), the results for a pure hexane monolayer are presented in very light gray dots, to help aid the comparison of the monolayer with impurities to one without.

The final order parameter, *OPtilt*, is a measure of the orientation of the molecules in the *z*-direction with respect to the plane parallel to the substrate and is defined as:

$$OPtilt = \frac{1}{2N_m}\left\langle \sum_{i=1}^{N_m}(3\cos^2\theta_i - 1)\right\rangle,$$

where $\theta_i$ is the angle that the plane extending about the long-axis of molecule (*i*) makes with the axis normal to the substrate. *OPtilt* is the thermal average of a Legendre polynomial ($P_2$), and takes a value of –0.5 if the long axis of each hexane molecule is parallel to the (*x,y*) plane. Figure 8 gives the average value of *OPtilt*, averaged over all 18 identical temperature points, as a function of the mole percent, for each type of impurity in the monolayer.

In addition to order parameters and energies, the atomic height distribution, *P(z)*, is monitored throughout all simulations and represents the probability of finding a particular pseudoatom at a height *z* above the substrate. This quantity is especially useful because it not only gives information regarding layer promotion in the intermediate and fluid phases, but also gives information regarding in-plane rolling of the molecules with the plane of their short-axis normal to the substrate. If all molecules are rolled in such a way, there will be two distinct peaks of equal amplitude present at two different values of *z*, separated by a distance that corresponds to the *z* projection of the pseudoatom bond length. If all molecules are rolled flat in the plane of the substrate, there will be one distinct peak representing a single value of *z* for all pseudoatoms. Figure 9 uses the atomic height to define whether an adatom is promoted to the 2$^{nd}$ layer (the adatom is considered "promoted" if it is greater than 5Å from the graphite surface), and is a plot of the total percent of impurities and the total percent of hexanes promoted to the 2$^{nd}$ layer as a function of the impurity chain length for monolayers containing 15% impurities at a temperature point well into the intermediate phase. In addition, figures 10 shows *P(z)* as a function of impurity chain length for 3%, 5%, 10%, and 15% impurities at *T*=115K.

## V. DISCUSSION

This section will be separated into three individual subsections. The first subsection will describe the observations of this work for hexane monolayers with impurities in the monolayer having a chain length smaller than that of hexane, the second subsection will present a discussion for impurities in the hexane monolayer that have a greater chain length than that of hexane, and the final subsection will present a comparison of pentane and heptane impurities, which provides an intriguing contrast of possibly the most important observations in the first two sections. Although there is no particular difference in how these are actually simulated, we believe that the chain length of the impurities plays a significant role on how the results should be interpreted, and such a division of this section provides the two emerging pictures of this system that we think is most useful to communicate to the reader.

### A. IMPURITIES WITH $n < 6$

Impurities that are studied in this work with a chain length smaller than the hexane molecules (whose monolayer these impurities are adsorbed into) consist of propane (C3) and pentane (C5). As mentioned previously, a solution containing these two molecules would be expected to (especially in the case of pentane, whose alkyl length is very similar to hexane) adsorb into the monolayer in some small concentration (due to preferential adsorption), depending on the amount of the respective impurity present in the initial solution. From figure 2, one can notice from *OPcom* and $<U_1>$, that the only monolayer containing impurities smaller than hexane at 5% mol concentration and exhibiting commensurability with the substrate comparable to the hexane monolayer is that containing the C5 impurities. In fact, these quantities show that the C5 impurities indicate very similar patterns in the center-of-mass and atomistic commensurability as compared to the pure hexane monolayer, except for a slightly early phase transition into an incommensurate phase that is observed compared with the monolayer containing C5 impurities. However, such is not the case for the C3 impurities. Although the monolayer with C3 impurities shows a glint of commensurability with the substrate until ca. *T*=120K, the commensurability that is achieved is of no comparison to the pure hexane monolayer, or even the monolayer containing the pentane impurities. This effect that involves the loss of commensurability for the monolayer containing the C3 impurities may be understood by figures 9 and 10. In figure 9, the average percent of molecules being promoted to the bilayer at a temperature point in the intermediate phase (for 15% C3 impurities) is presented in terms of impurity molecules and hexane molecules. One notices that there is ca. 40% of the C3 molecules being promoted to the 2$^{nd}$ layer, whereas there are effectively no hexanes being promoted. This is not surprising, since simulation of a hexane monolayer with C3 impurities presents a "less than monolayer" coverage, but it does not present space in the conventional way used to simulate submonolayer hexane in [24] (with a uniaxial expansion). However, if the C3 molecules are undergoing phase separation (and hence, demixing) to the bilayer, then this produces large vacancies in the monolayer. Previous work over the pure hexane monolayer [11] has shown that the phase transition into the intermediate phase is facilitated by a significant number of hexane molecules being promoted to the $2^{nd}$ layer, creating vacancies in the monolayer. Thus, it is not surprising that the intermediate phase transition temperatures presented in table I for monolayers containing C3 impurities are the lowest of all the impurities studied in this work. This is also supported by analysis of figure 10 in comparison to the transition temperatures presented in table I. Figure 10 gives the atomic height distribution for various impurity types vs. the mole percent of the impurities in the monolayer at $T$=110K. The choice of this temperature for this distribution gives a very interesting comparison between the intermediate phase transition temperatures reported in table 1 for C3 impurities at 3% and 5%. One notices that at 3% impurities, the intermediate phase transition temperature is noted as $T_I$=105K. Comparing this to figure 10, one observes a significant amount of promotion to the bilayer at $T$=110K for this impurity composition. However, the phase transition temperature into the intermediate phase for 5% C3 impurities is given as $T_I$=112K, slightly higher than $T$=110K as shown in figure 10. However, upon inspection of figure 10, one notices that there is very little observed layer promotion to the bilayer in this case. This could emphasize the early phase transition into the intermediate phase as being a result of phase separation of the C3 molecules to the bilayer. However, it should be noted that from figure 3 and table I, that even 3% impurities of C3 has a major effect on the phase behavior of the monolayer.

In addition to giving information regarding the phase transition from a solid herringbone phase into an intermediate phase, figure 2 also indicates the behavior of the monolayer during the intermediate phase, as well as at the melting transition. From *OPnem* in figure 2, one observes the heightening of *OPnem* from the ca. *OPnem*=0.5 that is characteristic of the phase transition into the intermediate phase and the vanishing of *OPnem* at higher temperatures which is indicative of the loss of structural order, and hence melting. In agreement with the phase behavior observed previously for the intermediate phase in [11], one can notice that monolayers with both C3 and C5 impurities give a heightened value of *OPnem* with the maximum being ca. 0.7 (for C3) and 0.8 (for C5). This indicates that the presence of the impurities has little effect on this observed phase, but in the case of C3 impurities, the phase transition occurs at a significantly lower temperature. Also, one can notice from the average pair interaction energy ,$<U_{LJ}>$, in figure 2 that monolayers with pentane impurities seem to give the "U" shaped minimum in the interaction energy at the intermediate phase transition consistent with the general trend observed for the pure hexane monolayer. Although the features for this observation in the monolayer with C3 impurities are much more broadened, a slight lowering of the interaction energy in the intermediate phase is still observed.

Despite the differences at the intermediate phase transition for monolayers with C3 and C5 impurities, there seems to be very little difference present at the melting transition (at 5% C3 and C5 impurities in the monolayer). This is evident by figure 2 with *OPnem* indicating that the loss of order in the monolayer occurs largely in the same temperature region for both monolayers. This is quantified in table I, as one sees that the melting temperatures at 5% are largely the same (with a 3K difference, which is the smallest proposed uncertainty for either case). However, further inspection of larger amounts of impurities in the monolayer (10% and 15%) indicates that monolayers with C5 impurities seem to indicate a fairly consistent melting temperature over all mole fractions studied, however, monolayers with C3 impurities seem to significantly decrease in melting temperature as the number of C3 impurities in the monolayer is increased. Although this difference in the melting temperature can only raise speculation, we believe that this is due to the larger space reduction that exists in a hexane monolayer with an increased number of C3 impurities (and hence more phase separation). From figure 9 one observes ca. 40% (at 15% impurities in the intermediate phase) of the C3 molecules, on average, are promoted to the bilayer. In the case of 3-5% impurities of C3 in the monolayer, the phase separation does not seem to create enough in-plane space to substantially affect the melting temperature. However, as the number of C3 impurities is increased, more C3 impurities means that (if the 40% is consistent over all coverages) there are more C3 molecules being promoted to the bilayer, and hence more space created in the monolayer. Previous experimental work over submonolayer hexane [19] indicates that the melting temperature decreases slightly at lower coverages, so it is a possibility that the space reduction created by the layer promotion of the C3 molecules in concentration of 10% and 15% in the monolayer could create enough space to affect the melting transition, where as smaller concentrations do not significantly change the melting temperature (as is observed in chapter 3).

### B. Impurities with $n>6$

As mentioned previously, impurities in the monolayer with a longer chain length than that of the hexane monolayer in which they are adsorbed into bring a very unique interpretation to this study that is important for understanding the effects of defects in a monolayer based upon uncertainty in the purity of the solution used to prepare the sample (due to preferential adsorption). The impurities that are studied in this work that involve a chain length greater than hexane are heptane (C7), nonane (C9), and undecane (C11). One can visually inspect the effects of such impurities on the solid herringbone phase in figure 1, for impurities with a mole fraction of 0.1 at $T$=110K. In general, one can observe an increasing distortion in the solid monolayer as the chain length of the impurity molecules increase from that of C7. This already justifies that there are some significant effects that arise in such a case of longer alkane impurities.

This is further emphasized by figures 5 and 6. Each figure contains the same quantities that were presented in figure 2, with figure 5 representing 3% impurities and figure 6 representing 5% impurities in the monolayer. By analyzing the measures of atomic and center-of-mass commensurability in both figures, one notices that the only case that compares to the pure hexane monolayer is that of 3% C7 impurities, which seems to also predict an early transition into an incommensurate phase. The other impurities at 3% and 5% seem to indicate minimal commensurability (max $OPcom$ is ca. 0.2 in both cases), indicating very significant effects on the monolayer commensurability resulting from these longer alkane impurities. Inspection of table I indicates that for the longer chain alkane impurities (except for 3% C7 impurities), the intermediate phase transition occurs at temperatures at or below ca. $T$=120K in all cases. Comparing this to observations with monolayers containing C5 impurities, one notices that the intermediate phase transition in all cases of C5 impurities that are studied occurs above $T$=120K. This emphasizes the effects of chain length on the intermediate phase transition, and this will be further discussed in the next section through a comparison of monolayers containing C5 impurities to those containing C7 impurities.

      Further analysis of the phase transition temperatures in table I indicates the trends observed in the melting behavior of the longest impurity, C11, are very similar to those observed for the shortest impurity, C3, in that they are significantly lower than that observed for the pure hexane monolayer. However, in contrast to monolayers with C3 impurities, monolayers with C11 impurities show a gradual decrease in the melting temperature as the number of impurity molecules is increased (as opposed to the sharp decline in melting temperature observed for the C3 impurities). Also, analogous to the phase separation behavior observed in monolayers with C3 impurities, figure 9 indicates that the monolayers with C11 impurities exhibit a significant amount of phase separation, but in a different way. In general, since the chain length mismatch between C11 and C6 is one of significance, one expects that preferential adsorption requires that due to the increased number of hydrogen atoms per surface area in the longer alkane, it will adsorb onto the surface first. Figure 9 seems to indicate this behavior from another perspective, which is that of the significant promotion of the hexane molecules, and the very small promotion of the longer impurity molecules. This is indicative of a type of phase separation of the C6 molecules from the impurity molecules, and is very interesting in nature. However, it should also be noted that in the cases of C3 and C11 impurities, these are largely studied in this work to further analyze the evolving trends associated with the increasing/decreasing chain length of impurities in the monolayer. In fact, with such a significant chain length mismatch between the two adsorbates (hexane and C11), one may expect that a solid herringbone phase would not accommodate such size of impurities, and that the effects would be more serious than the case of the molecule being "placed" in the monolayer. This may be what simulations are indicating in figure 8 and 9, as well as table I, which all seem to suggest some significantly different behavior in the tilting, layer promotion, and melting transition behavior for these monolayers as compared to other impurities studied. However, it is beyond the scope of this work (and is

generally experimentally unexplored) to study the effects of crystallization of a solid monolayer under such conditions. However, it marks the beauty of simulations to include such systems to further understand how they compliment those systems in which we do believe are important toward understanding such a real effect of impurities in a monolayer.

Therefore, this section demonstrates that there are, in fact, significant effects that come about due to the adsorption of impurities with a chain length longer than that of a hexane molecule. The intermediate phase transition temperatures are significantly lowered, and although the melting transition behavior is not significant affected, there is some observable difference that is introduced by these impurities.

### C. Impurities with $n=5$ and $n=7$

This last section involves a very unique comparison of two impurities, pentane (C5) and heptane (C7), that are very similar in their chain length with respect to the hexane monolayer in which they are adsorbed into, but introduce significant differences in how the monolayer is affected by their presence. In fact, this is easily seen by studying figure 4, where the herringbone order parameter is plotted for all compositions and temperatures that are simulated in this work. In general, one observes that for 3% and 5% impurities, the monolayer containing the C5 impurities seems to correspond relatively well to that of the pure hexane monolayer, whereas there is an obvious difference in the herringbone phase behavior in the monolayer containing the C7 impurities as compared to the pure hexane monolayer. In fact, by comparing the phase behavior of the monolayers containing C5 and C7 impurities at all compositions studied, one notices that the herringbone order in the monolayer containing the C5 impurities is *always* more closely associated with the pure hexane monolayer than the one containing the C7 impurities.

This comparison is a unique feature that separates the phase behavior of the monolayer containing these two similar impurities, and we propose that this is largely due to the symmetry of the sublattices of a solid herringbone phase. In general, in the case of C5 impurities, the pentane molecules can better "fit" into the spaces in the herringbone phase without a significant effect of distortion in the monolayer. However, such is not the case for the C7 impurities, as they are still very close in length to the hexane molecules, but due to the larger chain length of the C7 molecules, the ends (the methyl groups) of the molecule will protrude from the sides of the sublattices, causing a slight distortion of the solid herringbone phase, which is much more evident with more impurities in the monolayer. However, one notices that the monolayer with 3% C7 impurities (from figure 5) still achieves a significant amount of commensurability at low temperatures, and it must be due to the increased thermal energy (and hence, fluctuations) that cause an increased interaction between the protruding molecules and the neighboring sublattices. However, when one inspects the melting transition temperatures, one notices that they are almost identical for monolayers with C5 and C7 impurities, with up to 17 impurity molecules in the monolayer. This indicates the role of the intermediate phase (which is emphasized in [11]) which is to "regulate" the space in the monolayer as to minimize the interaction energy. In such a phase, where the molecule-molecule interaction dominates the phase behavior, a small increase/decrease in chain length with respect to this interaction does not have same effect as it has on the solid phase, because in the solid phase, the order is determined strictly by the symmetry introduced by the corrugation in the graphite surface. Thus, this section emphasizes the "sublattice protrusions" that seem to lower the intermediate phase transition temperatures for monolayers containing C7 impurities as compared to those containing C5 impurities, however, due to the nature of the interactions in the intermediate phase, the melting temperatures for both monolayers are almost identical with up to 15% impurities of C5 and C7 in the monolayer.

## VI. CONCLUSIONS

Therefore, the study presented in this work provides three distinct conclusions about the phase behavior for monolayers of hexane that involve odd-alkane impurities. (*i*) A monolayer containing C5 impurities gives the best correspondence to that observed for a pure hexane monolayer, as compared to any other type of odd-alkane impurity studied. In terms of a possible interpretation of this result, this could indicate that in a binary mixture of pentane and hexane, with a small amount of pentane adsorbed into the hexane monolayer, that the effects of the pentane on the phase behavior of the monolayer will not tend to give significantly different results regarding the phase transitions and phase behavior. (*ii*) Due to "sublattice protrusions," impurities in the monolayer with a chain length greater than hexane will have a "profound consequence" on the solid herringbone phase, leading to a significantly early phase transition to a phase where the molecule-molecule interaction is dominant. However, with only few of these impurities in the monolayer, the affect on the melting transition is not as apparent, with the only significant differences occurring for C11 (whose chain length is quite a bit longer than that of hexane) and above 10% of C9 impurities. Since the longer molecule is expected to adsorb onto the surface either prior to or in a greater abundance than the hexane molecules, this observation provides valuable insight into the ramifications of utilizing even a slightly impure solution for such a process. (*iii*) Monolayers studied in this work with propane and undecane impurities seem to suggest that their behavior is related in that it doesn't correspond well (in any case) to that for the pure hexane monolayer. This type of behavior (including the significant amount of layer promotion and phase separation) leaves one to wonder whether these systems are representative of a "real" system occurring in nature. However, they still introduce a sense of "completeness" to the study of the other odd-alkane impurities, and further support that the simulations presented in this work show indications of both phase separation and preferential adsorption, which is important due to the recent difficulty in observing such behavior in simulations for alkanes with mismatched chain lengths [25].

## ACKNOWLEDGEMENTS

The authors are both indebted to Paul Gray and the UNI CNS for use of generous usage of CPU time which was very valuable to the completion of this work.

# REFERENCES

1. J. Krim, J. Suzanne, H. Shechter, R. Wang and H. Taub, Surf. Sci. **162**, 446 (1985).
2. J. C. Newton, Ph.D. Dissertation, University of Missouri-Columbia, 1989.
3. H. Taub, in *NATO Advanced Study Institutes, Series C: Mathematical and Physical Sciences*, edited by G. J. Long and F. Grandjean (Kluwer, Dordrecht, 1988), Vol. 228, pp. 467-497.
4. T. Arnold, R.K. Thomas, M.A.. Castro, S.M. Clarke, L. Messe, and A. Inaba, Phys. Chem. Chem. Phys. **4**, 345 (2002).
5. F. Y. Hansen and H. Taub, Phys. Rev. Lett. **69**, 652 (1992).
6. F. Y. Hansen, J.C. Newton and H. Taub, J. Chem. Phys. **98**, 4128 (1993).
7. E. Velasco and G. H. Peters, J. Chem. Phys. **102**, 1098 (1995).
8. G. H. Peters and D. J. Tildesley, Langmuir **12**, 1557 (1996).
9. G. H. Peters, Surf. Sci. **347**, 169 (1996).
10. K.W. Herwig, Z.Wu, P. Dai, H. Taub, and F.Y. Hansen, J. Chem. Phys. **107**, 5186 (1997).
11. M. W. Roth, C. L. Pint and C. Wexler, Phys. Rev. B **71**, 155427 (2005).
12. B. Matthies, Ph.D. Dissertation, University of Missouri-Columbia, 1999.
13. T. Arnold, C.C. Dong, R.K. Thomas, M.A. Castro, A. Perdigon, S.M. Clarke, and A. Inaba, Phys. Chem. Chem. Phys. **4**, 3430 (2002).
14. A.J. Groszek, Proc. Roy. Soc. London A **314**, 473 (1970).
15. M.A. Castro, S.M. Clarke, A. Inaba, T. Arnold, and R.K. Thomas, J. Phys. Chem. B **102**, 10528 (1998)
16. M.A. Castro, S.M. Clarke, A. Inaba, R.K. Thomas, and T. Arnold, Phys. Chem. Chem. Phys. **3**, 3774 (2001).
17. A. Inaba, S.M. Clarke, T. Arnold, and R.K. Thomas, Chem. Phys. Lett. **352**, 57 (2002).
18. E. Sloutskin, X.Z. Wu, T.B. Peterson, O. Gang, B.M. Ocko, E.B. Sirota, and M.Deutsch, Phys. Rev. E **68**, 31605 (2003).
19. E.P. Gilbert, P.A. Reynolds, P. Thiyagrajan, D.G. Wozniak, and J.W. White, Phys. Chem. Chem. Phys. **1**, 2715 (1999).
20. W.A. Steele, Surf. Sci. **36**, 317 (1973).
21. M. G. Martin and J. I. Siepmann, J. Phys. Chem. **102**, 2569 (1998)
22. P.Padilla and S. Toxværd, J. Chem. Phys. **94** 5650 (1991).
23. M. P. Allen , D. J. Tildesley, *Computer simulation of liquids* (Clarendon Press, New York, 1988).
24. C.L. Pint, M.W. Roth, and C. Wexler, Phys. Rev. B (submitted).
25. P. Smith, R.M. Lynden-Bell, and W. Smith, Mol. Phys. **98**, 255 (2000).

**TABLES AND FIGURES**

| Impurity type | 3% | | 5% | | 10% | | 15% | |
|---|---|---|---|---|---|---|---|---|
| | $T_1$ | $T_2$ | $T_1$ | $T_2$ | $T_1$ | $T_2$ | $T_1$ | $T_2$ |
| C3 (Propane) | 105±5 | 172 | 112 | 175±5 | < 100 | 158 | < 100 | 150±5 |
| C5 (Pentane) | 135±5 | 172 | 128 | 172 | 122 | 170±5 | 122 | 168 |
| C7 (Heptane) | 128 | 172 | 112 | 170±5 | 112 | 170±5 | 105±5 | 168 |
| C9 (Nonane) | 118 | 168 | 118 | 172 | 110±5 | 168 | 105±5 | 155±5 |
| C11 (undecane) | 115±5 | 165±5 | 120±5 | 160±5 | 105±5 | 158±7 | <100 | 155±5 |

**Table I.** Transition temperatures, $T_1$ and $T_2$, for 3%, 5%, 10%, and 15% impurities of propane, pentane, heptane, nonane, and undecane. For $T_1$, *OPher*, $<U_1$, $<U_{LJ}$, *OPcom*, and *OPnem* were used to quantify this transition, whereas for $T_2$, OPnem and $U_{LJ}$ were used. All temperatures have units of K, and an uncertainty of ± 3K unless otherwise stated.

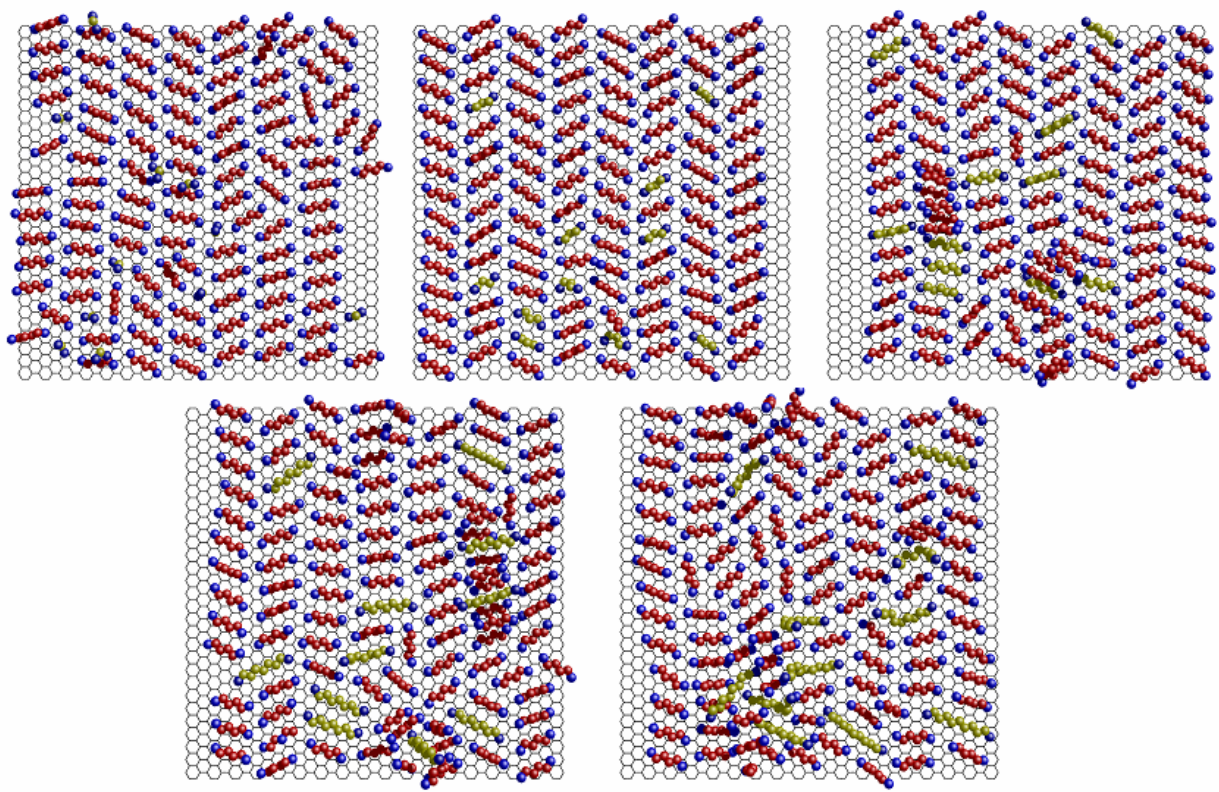

(Color Online) **Figure 1.** Snapshots of the hexane monolayer containing 10% impurities of propane (top, left), pentane (top, middle), heptane (top, right), nonane (bottom, left), and undecane (bottom, right). All snapshots are taken at $T$=110K, with hexane molecules red/blue and impurity molecules yellow/blue.

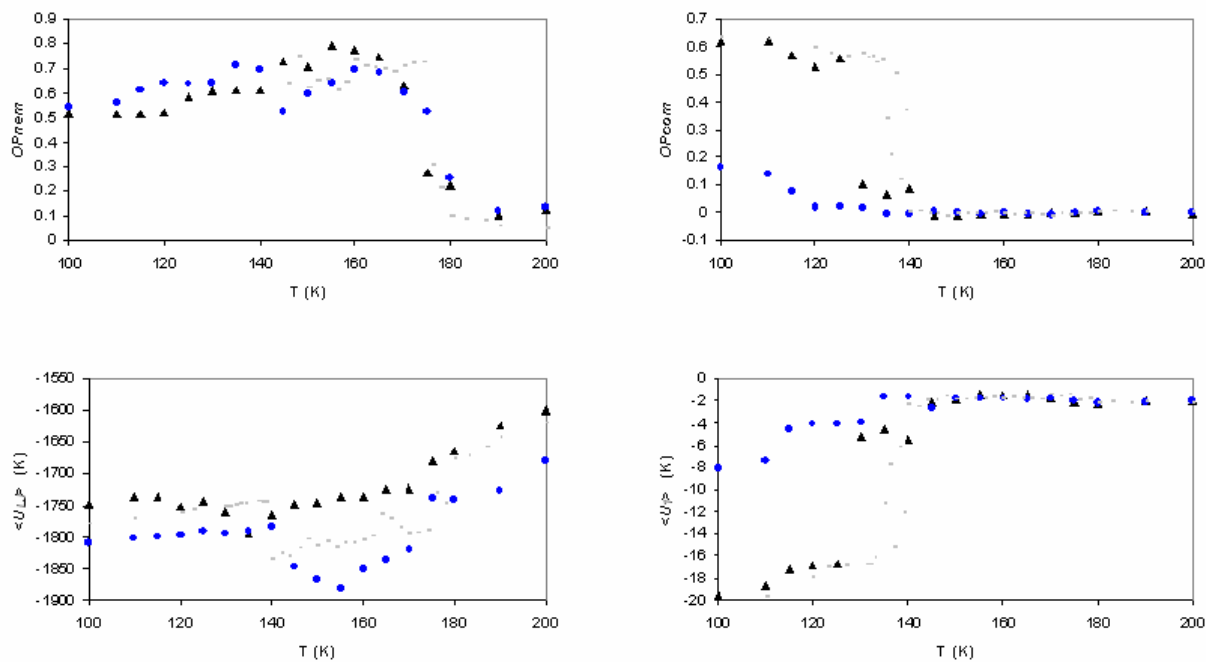

(Color Online) **Figure 2**. (Top) Order parameters *OPnem* (left) and *OPcom* (right), and (bottom) average interaction energy (left) and average corrugation potential energy (right) for 5% propane (blue circles) and pentane (black triangles) impurities in the hexane monolayer. The gray dots correspond to graphs for a pure hexane monolayer, for comparison.

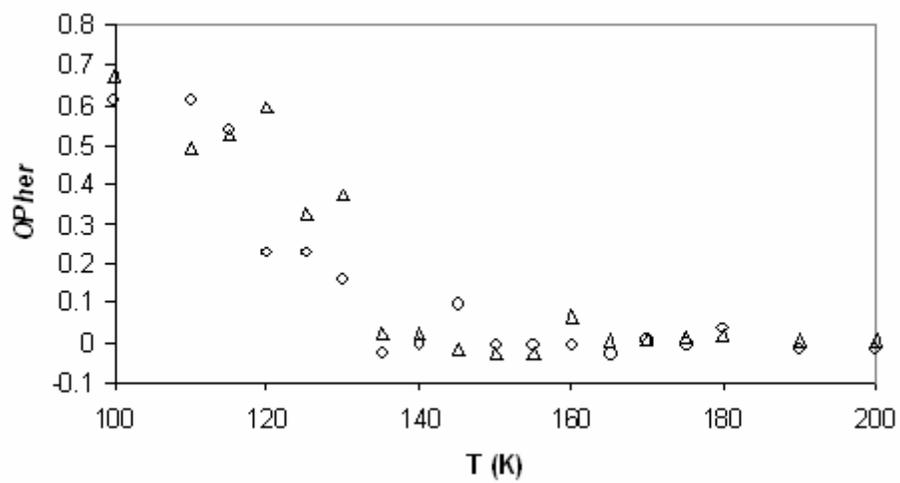

**Figure 3.** Temperature dependence of the herringbone order parameter for a monolayer containing 3% (circles) and 5% (triangles) impurities of propane.

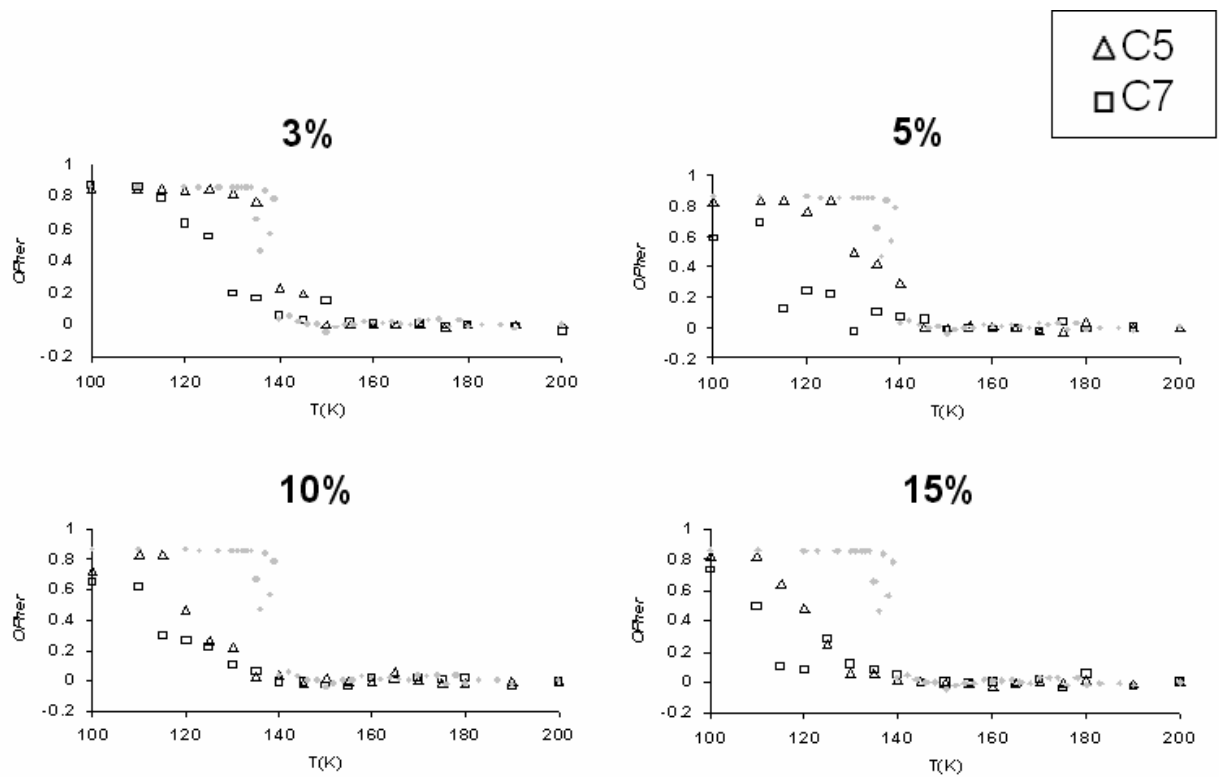

**Figure 4**. Temperature dependence of the herringbone order parameter, *OPher*, for C5 (triangles) and C7 (squares) adsorbed as impurities in the monolayer in labeled amounts. Again, the light gray points correspond to the temperature dependence of *OPher* for a pure hexane monolayer.

**Error!**

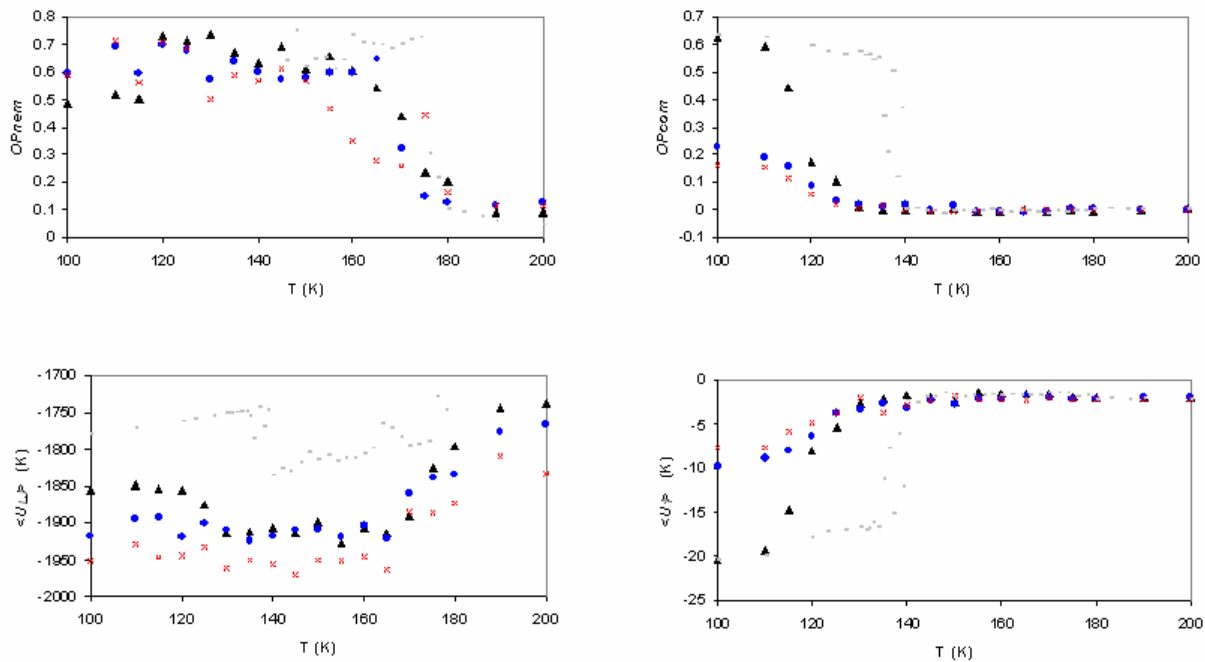

(Color Online) **Figure 5.** Temperature dependence of order parameters and energies (format the same as in figure 2) for monolayers with 3% impurities of heptane (black triangles), nonane (blue circles), and undecane (red x's). Again, the gray dots represent these quantities for a pure hexane monolayer.

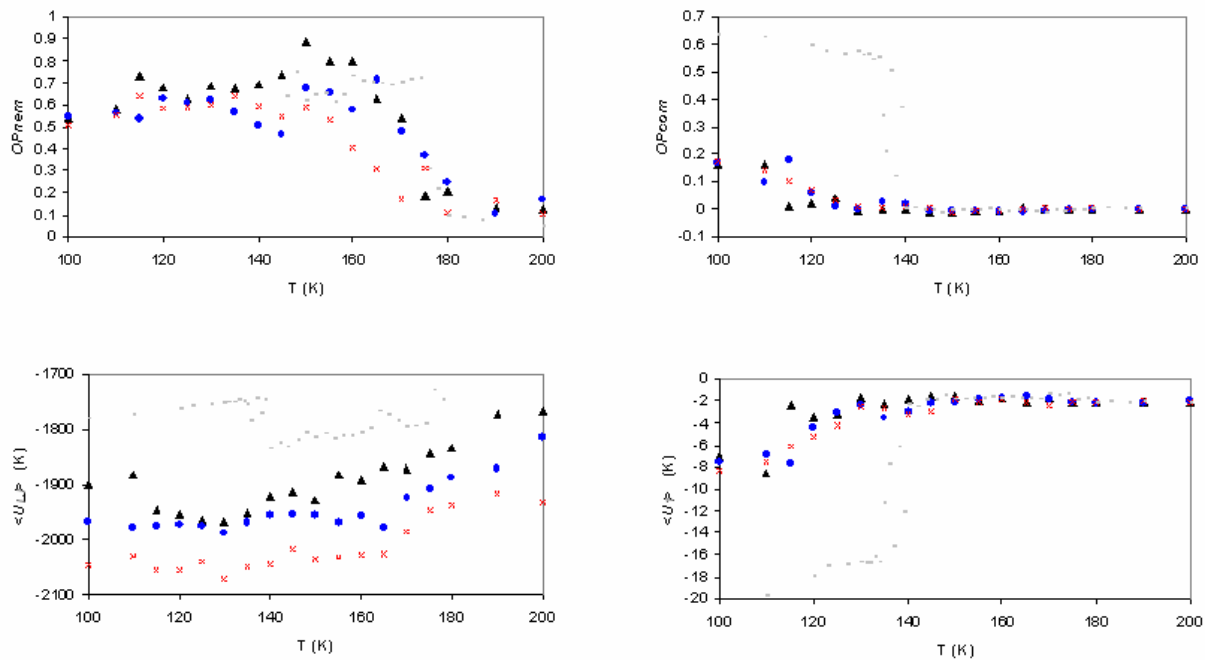

(Color Online) **Figure 6.** Temperature dependence of order parameters and energies (format the same as in figures 2 and 5) for monolayers containing 5% impurities of heptane, nonane, and undecane, each labeled in an identical fashion to that introduced in figure 7-5.

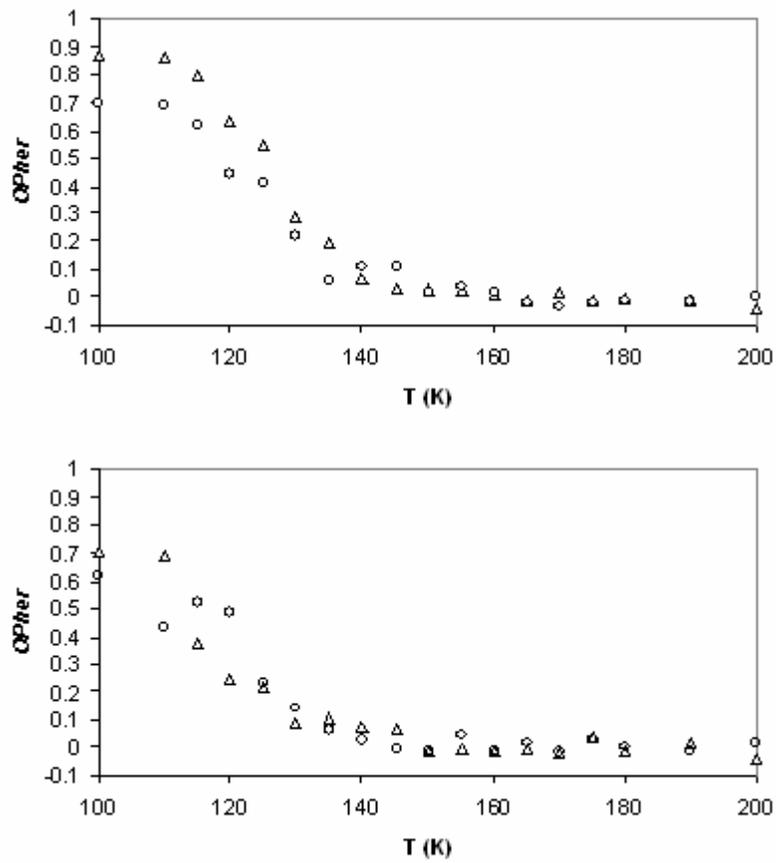

**Figure 7.** Temperature dependence of the herringbone order parameter, *OPher*, for monolayers of 3% (top) and 5% (bottom) impurities of heptane (triangles) and nonane (circles).

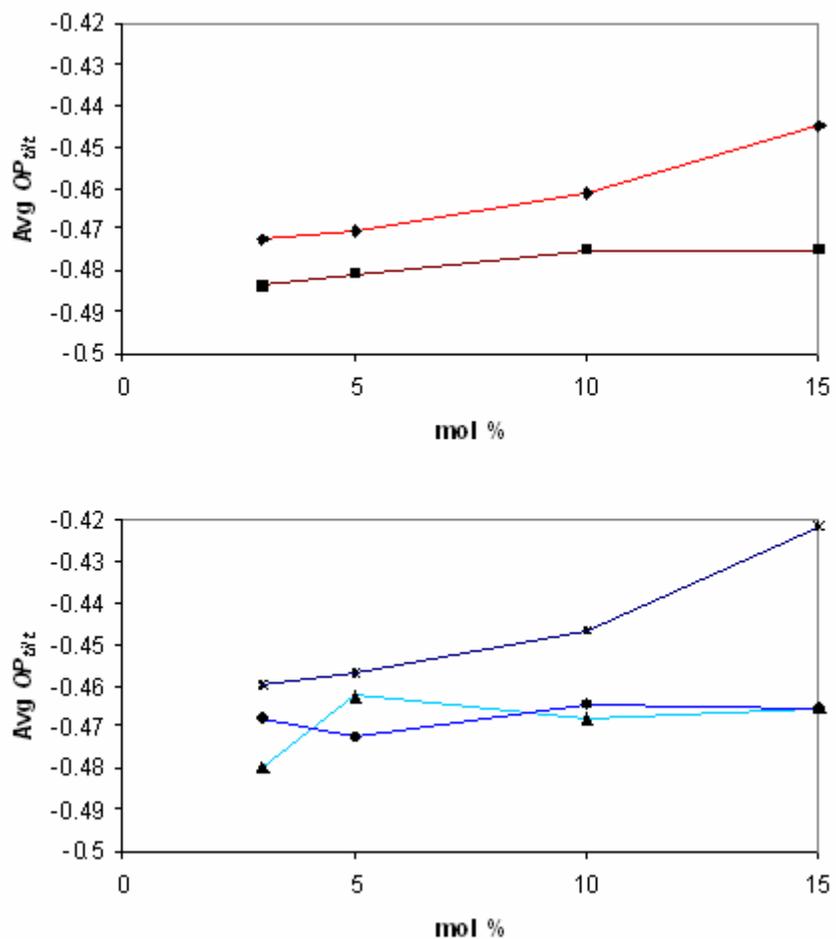

(Color Online)**Figure 8.** Average tilt order parameter (averaged over all temperature points) as a function of mole percent for alkane impurities (top) shorter (triangles for C3 and squares for C5) than hexane and (bottom) longer (triangles for C7, circles for C9, and stars for C11) than hexane. The color of the lines darken with increasing chain length.

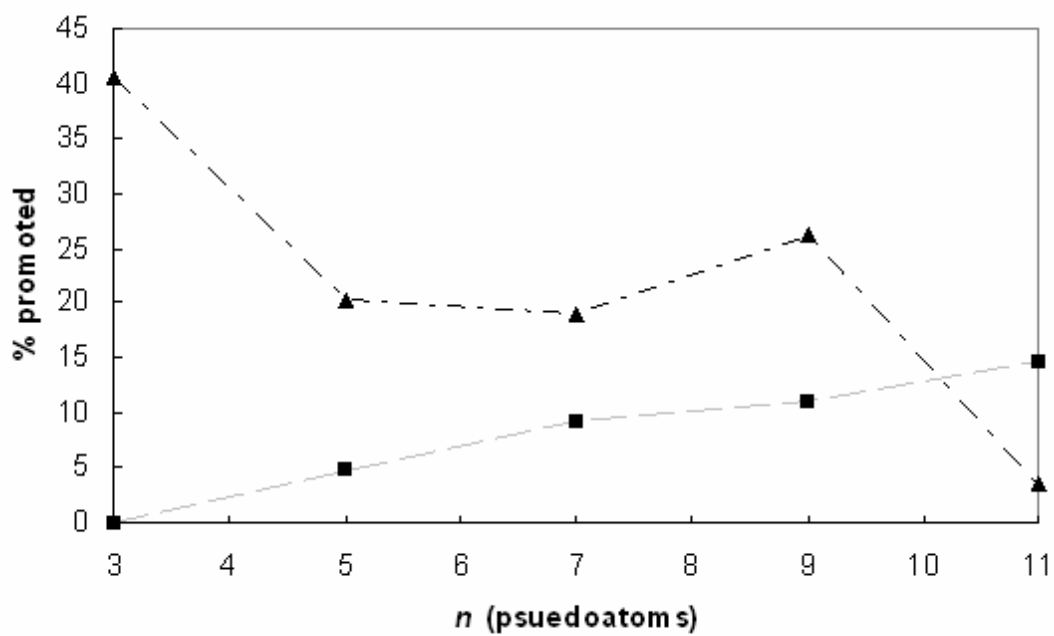

**Figure 9.** The percent of impurities (triangles) and hexane molecules (squares, bold) promoted to the bilayer at a characteristic temperature point well into the intermediate phase, plotted as a function of the number of psuedoatoms in the chain length of the impurities.

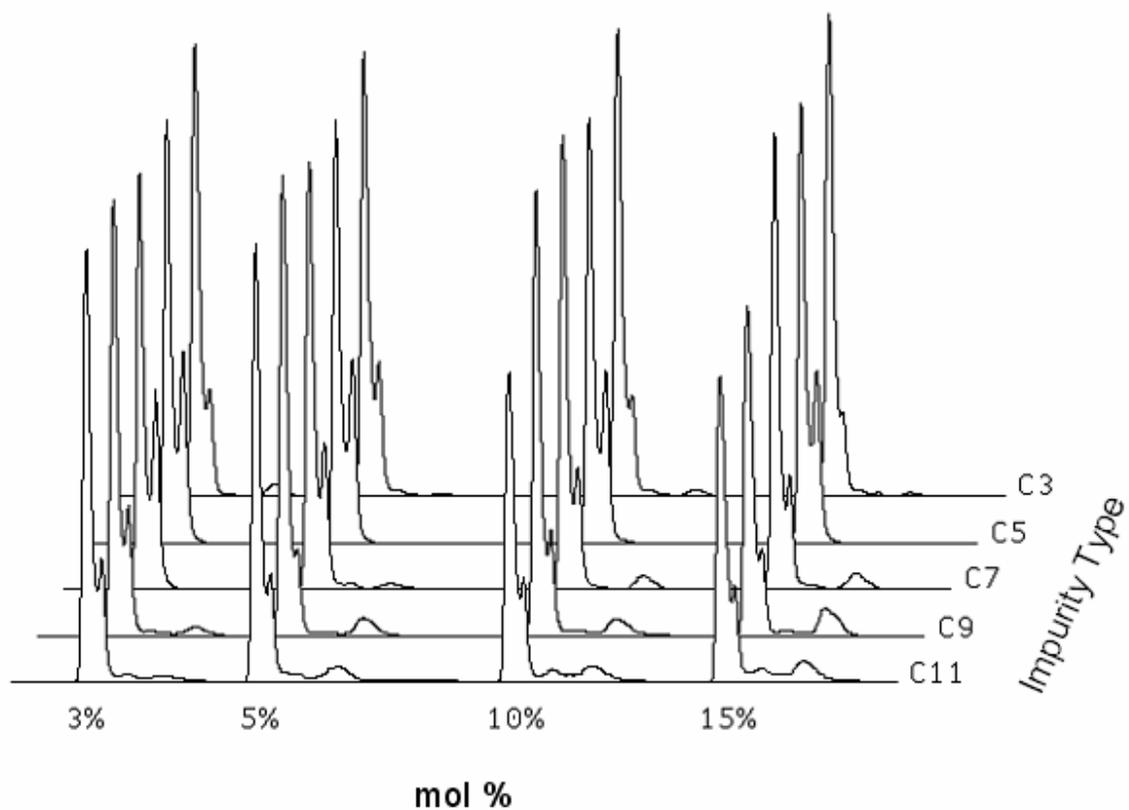

**Figure 10.** Atomic height distributions at $T$=110K for the simulated impurities and their corresponding mole percent. Impurity chain length increases from back to front, and the percent of impurities in the monolayer increases from left to right.